\documentclass[jcp, floatfix, nobibnotes, reprint, superscriptaddress]{revtex4-1}
\usepackage[english]{babel}
\usepackage[utf8]{inputenc}
\usepackage[T1]{fontenc}
\usepackage{calrsfs}
\usepackage{caption} 
\usepackage{amsmath}
\usepackage{mathalpha}
\usepackage{amssymb}
\usepackage{amsfonts}
\usepackage{graphicx}
\usepackage{subcaption}
\usepackage{dcolumn}
\usepackage{rotating}
\usepackage{relsize}
\usepackage{bm}
\usepackage{multirow}
\usepackage{bigints}
\usepackage[varg]{txfonts}
\usepackage{bbm}
\usepackage{listings}
\usepackage[hmargin=2.5cm, vmargin=4cm]{geometry}
\graphicspath{{figures/}}
\usepackage{url}
\usepackage{braket}
\usepackage{textcomp}
\usepackage{float}
\usepackage{bbold} 
\usepackage{comment}   
\usepackage{suffix}
\usepackage[dvipsnames]{xcolor} 
\usepackage[normalem]{ulem}

\newcommand{\papertitle}{Alchemical harmonic approximation based potential for iso-electronic diatomics: 
Foundational baseline for $\Delta$-machine learning}

\def\beq{\begin{equation}}
\def\eeq{\end{equation}}
\def\bea{\begin{eqnarray}}
\def\eea{\end{eqnarray}}
\def\brcl{\begin{array}{rcl}}
\def\bccl{\begin{array}{ccl}}
\def\blcl{\begin{array}{lcl}}
\def\err{\end{array}}

\begin{document}

\title{\papertitle}

\author{Simon León Krug}
\email{simonleon.krug@gmail.com}
\affiliation{Machine Learning Group, Technische Universität Berlin, 10587 Berlin Charlottenburg, Germany}

\author{Danish Khan}
\email{danishk.khan@mail.utoronto.ca}
\affiliation{Vector Institute for Artificial Intelligence, Toronto, ON M5S 1M1, Canada}
\affiliation{Department of Chemistry, University of Toronto, St. George campus, Toronto, ON M5S 3H6, Canada}

\author{O. Anatole von Lilienfeld}
\email{anatole.vonlilienfeld@utoronto.ca}
\affiliation{Machine Learning Group, Technische Universität Berlin, 10587 Berlin Charlottenburg, Germany}
\affiliation{Vector Institute for Artificial Intelligence, Toronto, ON M5S 1M1, Canada}
\affiliation{Department of Chemistry, University of Toronto, St. George campus, Toronto, ON M5S 3H6, Canada}
\affiliation{Berlin Institute for the Foundations of Learning and Data, 10587 Berlin Charlottenburg, Germany}
\affiliation{Acceleration Consortium, University of Toronto. 80 St George St, Toronto, ON M5S 3H6, Canada}
\affiliation{Department of Materials Science and Engineering, University of Toronto, St. George campus, Toronto, ON M5S 3E4, Canada}
\affiliation{Department of Physics, University of Toronto, St. George campus, Toronto, ON M5S 1A7, Canada}

\date{\today}

\begin{abstract}
We introduce the alchemical harmonic approximation (AHA) of the absolute electronic energy for charge-neutral iso-electronic diatomics at fixed interatomic distance~$d_0$. To account for variations in distance, we combine AHA with this Ansatz for the electronic binding potential,
$E(d)=(E_{u}-E_s) \left(\frac{E_c-E_s}{E_u-E_s}\right)^{\sqrt{d/d_0}}+E_s$, where $E_u,E_c,E_s$ correspond to the energies of united atom, calibration at $d_0$, and sum of infinitely separated atoms, respectively. Our model covers the two-dimensional electronic potential energy surface spanned by distances of~0.7 to 2.5~\AA and differences in nuclear charge from which only one single point (with elements of nuclear charge $Z_1,Z_2$ and distance $d_0$) is drawn to calibrate $E_c$. Using reference data from pbe0/cc-pVDZ, we present numerical evidence for the electronic ground-state of all neutral diatomics with 8, 10, 12, 14 electrons. We assess the validity of our model by comparison to legacy interatomic potentials (Harmonic oscillator, Lennard-Jones, and Morse) within the most relevant range of binding (0.7-2.5\,Å), and find comparable accuracy if restricted to single diatomics, and significantly better predictive power when extrapolating to the entire iso-electronic series. We also investigated $\Delta$-learning of the electronic absolute energy using our model as baseline. This baseline model results in a systematic improvement, effectively reducing training data needs for reaching chemical accuracy by up to an order of magnitude from $\sim$1000 to $\sim$100. By contrast, using AHA+Morse as a baseline hardly leads to any improvement, and sometimes even deteriorates the predictive power. Inferring the energy of unseen CO converges to a prediction error of $\sim$0.1 Ha in direct learning, and $\sim$3 Ha with our baseline.\footnote{Previous versions including the published paper (JCP 162, 044101 (2025)) report 0.04 Ha. This version fixes only that typo, nothing else.}
\end{abstract}

\maketitle

\section{Introduction}
Quantum mechanics underpins our ability to predict electronic, optical, and thermal properties with high fidelity, essential for understanding chemical space or designing materials with specific functionalities~\cite{ceder1998predicting}. Unfortunately, numerically solving meaningful approximations to the electronic Schr\"odinger equation on-the-fly for each and every material remains a computationally prohibitive challenge.
Significant acceleration can be achieved via machine learning (ML) based inference which can remove the need for its solution~\cite{rupp_muller_lilienfeld_2012,Elpasolite_2016,QMLessayAnatole}, or enhance the numerical methods~\cite{apbe0} commonly employed.
Although practical and more universally applicable by now, e.g. in the form of fragment based building blocks for quantum machine learning~\cite{Amons,huang2021abinitio}, or "general-purpose" models of force field potentials trained on diverse data~\cite{batatia2024foundationmodelatomisticmaterials},
freely sampling chemical space is still hampered due
 to ML models' inherent interpolative nature which 
 lacks the universality of the Schr\"odinger equation.
Quantum alchemy based techniques by contrast provide an interesting alternative, physically motivated scheme that exploits perturbation theory to bypass explicit solutions across different composition, i.e.~for new systems~\cite{AlchemyAlisa_2016,Samuel-JCP2016,Yasmine-JCP2017,Samuel2018bandgaps,rudorff2020_alchem_chirality, shiraogawa2023optimization,krug_generalAIT}. 
In contrast to machine learning, typically only one or very few solutions are required for calibration.
While less commonly deployed than their ML counterpart, quantum alchemical perturbation~\cite{von_Rudorff_2020} based methods have become universally applicable across chemical space, e.g.~Refs.~\onlinecite{,lilienfeld_tuckerman,weigend_firstalchemy,michael_hammett_JACS,balawender_2019,Alchemy_bindingenergies,williams-noonan_2018,Samuel2018bandgaps, rudorff2021arbitrarily}, and have even already been used to calculate meaningful baselines for $\Delta$-machine learning models of catalytic activity~\cite{griego2020machine}.

In this work we focus on alchemically motivated approximate interpolations of the entire series of nuclear transmutations in diatomics possible for a given number of electrons. 
Motivated by the concavity of the electronic potential in any alchemical changes~\cite{anatole-ijqc2013}, we have investigated low order even polynomials, i.e.~the alchemical harmonic approximation (AHA) and the alchemical quartic approximation (AQuA). 
In order to also account for the impact of interatomic distances, a novel potential has been developed. 
Due to its interpolative nature, our joint universal potential model of alchemical {\em and} geometrical changes goes beyond our recent preceding efforts to estimate geometry changes due to compositional changes~\cite{domenichini_2022_al_geom_relax} via perturbation theory and using the mixed Hessian involving nuclear charges and atomic positions (`alchemical forces')~\cite{fias_2019}.
Finding such models, i.e. quantitative descriptions of the interatomic potential for different nuclear charges, is an ongoing quest in different energies regimes, e.g. for van der Waals bonds \cite{khabibrakhmanov2023universal,matej_2024_QDO}. 
As many-body dispersion interactions are typically dominating the interatomic long-range regime,  this work focuses on the short-range distances governed by covalent binding.

\section{Theory}

\subsection{Alchemy}

Starting with Alchemical Perturbation Density Functional Theory (APDFT)~\cite{von_Rudorff_2020}, one considers any two iso-electronic systems $\hat{H}_A$, $\hat{H}_B$. For relative statements about their total energies~$U_A, U_B$, connect them with a parameter~$\lambda$ such that $\hat{H}(\lambda) = \hat{H}_A (1-\lambda)+ \hat{H}_B \lambda$. Then, the difference in energies~$\Delta U$ can be obtained from a Taylor-expansion of $\bra{\Psi} \hat{H}(\lambda) \ket{\Psi}$ w.r.t.~$\lambda$ at a distance of~$\Delta \lambda = 1$; as we are only interested in the electronic energy differences~$\Delta E$, we omit the difference in Coulomb repulsion~$\Delta E^{\text{NN}}$ for brevity's sake:
\begin{align}
    \label{eq:alchemical_expansion_lambda}
	\Delta E := E_B - E_A &= \sum_{p = 1}^{\infty} \frac{1}{p!} \frac{\partial^p E(\lambda)}{\partial \lambda^p} \bigg\vert_{\lambda = 0} 
\end{align}

Previous work has shown the convergence behavior of this series\cite{rudorff2021arbitrarily}, as well as its satisfactory accuracy, not just for symmetric systems like diatomics where every odd-numbered term of Eq.~\ref{eq:alchemical_expansion_lambda} vanishes\cite{vonlilienfeld2023_1st_order, rudorff2020_alchem_chirality}, but for truncation after its second order\cite{fias_2019}.

Given the strict condition that the $E$ is concave in any $\lambda$~\cite{anatole-ijqc2013}, we begin by assuming the shape of a parabola, and we seek suitable conditions to determine its three degrees of freedom. We are not bound to truncate Eq.~\ref{eq:alchemical_expansion_lambda} after the second order, but testing revealed numerical instabilities when considering higher orders. 
Also note that this quadratic Ansatz attempts to directly model the expectation value of the electronic Schr\"odinger equation,  in line with our recently published model of the energy of the free atom~\cite{krug2024energies}; and is to be contrasted with previous attempts to replace alchemical linear interpolations of the Hamiltonian by quadratic interpolations~\cite{accurate_abinitio}.

\subsection{The alchemical harmonic approximation (AHA)}
\label{sec:AHA}

\begin{figure}
    \centering
    \includegraphics[width=\linewidth]{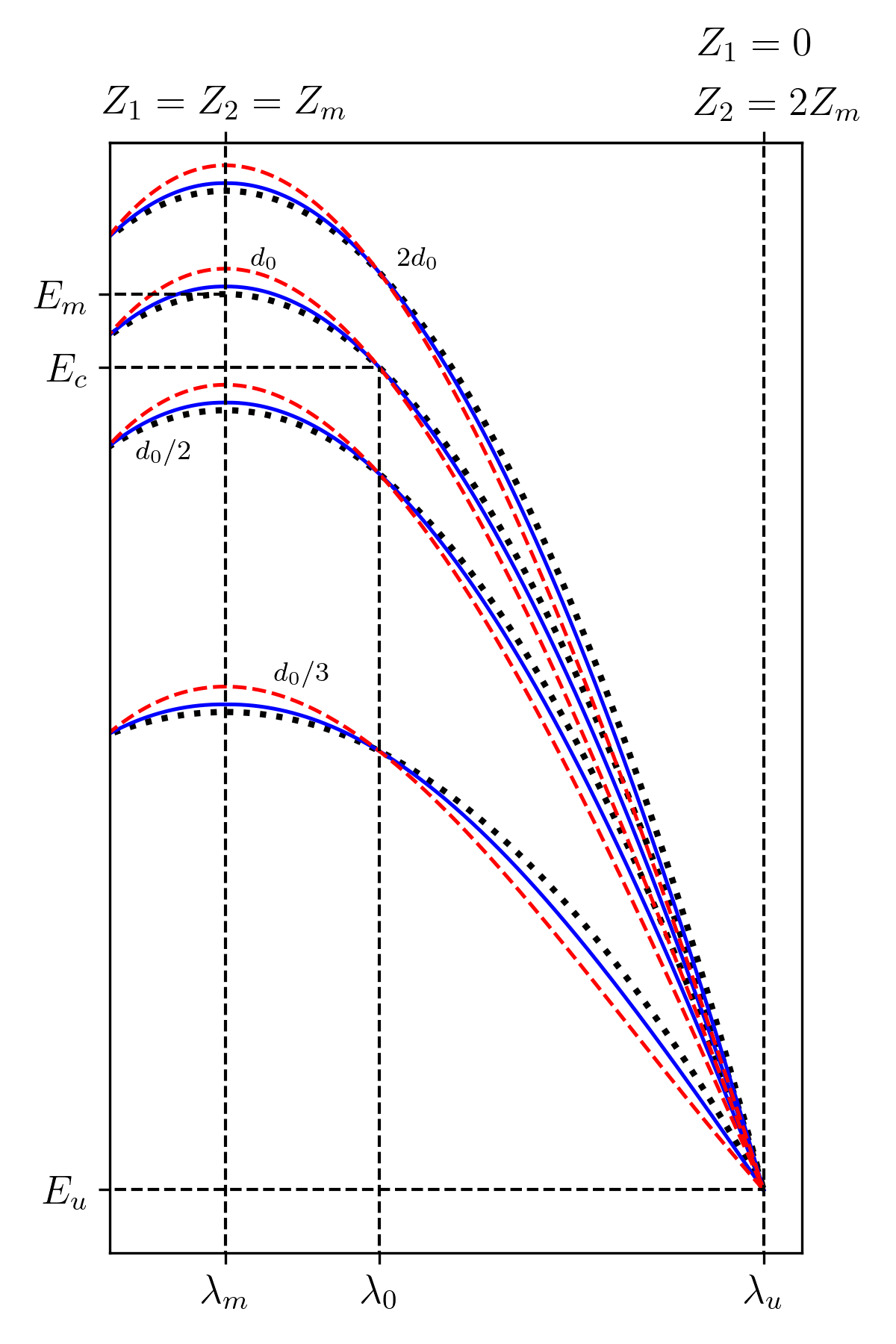}
    \caption{Schematic of reference (dashed red), AHA (dotted black), AQuA (solid blue) model of the absolute electronic energy for any iso-electronic diatomic series. Concave functions in nuclear charge differences~$\lambda$ are shown for four different interatomic distances~$d \in \{ 2d_0, d_0, d_0/2, d_0/3\}$. 
    For $d \rightarrow \infty$, the function would connect the energies of the free atoms, $E_s$. For $d \rightarrow 0$, the function converges to the energy of the united atom, $E_u$  where~$Z_1=0$ and~$Z_2=2Z_m$. Further annotations correspond to 
    the calibration point, $E_c = E(\lambda_0, d_0)$, and
    the homo-nuclear diatomic~($Z_1=Z_2=Z_m$) with its energy ridge at $E_m = E(\lambda_m)$.
    }
    \label{fig:cartoon_parabola}
\end{figure}

Consider a diatomic molecule with nuclear charges $Z_1, Z_2$, fixed interatomic distance $d_0$ and resulting electronic energy $E_c$.
To describe the entire series of iso-electronic diatomics, let us express changes in nuclear composition by the centered parameter $(\lambda - \lambda_m)$.
As we can pick the offset arbitrarily, let $\lambda_m = 0$; then $\lambda = (Z_2 - Z_1)/2 $ denotes the deviation from the symmetric diatomic of the considered series.
Now expand the energy of diatomics in powers of $\lambda$ up to and including second order (cf.~Fig.~\ref{fig:cartoon_parabola}).

To extend the energy prediction of the AHA from mere iso-electronic changes in nuclear composition to changes in distance $d$, we seek to construct the parabola of the AHA such that only one of the three necessary constraints exhibits $d$-dependency. For this, pick the endpoints of the energy parabola (where $\lambda = \pm (Z_1 + Z_2)/2 =: \pm \lambda_u$), i.e. the two united atoms ($Z_u := Z_1+Z_2$) with energy $E_u$ which is independent of  interatomic distance. 
The third, remaining constraint is given by a point in between, $E(\lambda_0, d)$, which \textit{will} change with varying distance. Our calibration point can be found at $d_0$, i.e. $E_0(d_0) = E_c$.
\begin{align}
    \label{eq:AHA_constraint_E_u}
    E(\pm \lambda_u) &=: E_u \\
    \label{eq:AHA_contraint_E_d}
    E(\lambda_0, d) &=: E_0(d)
\end{align}
Note, that both $E_u$ and $E_c$ must be in the same electronic state, e.g. computing $E_c$ at a small distance via quantum chemistry software will prove difficult as most methods implicitly assume the Born-Oppenheimer approximation, neglecting nuclear dynamics. Consequently, care must be taken, both with the calibration distance~$d_0$ and the corresponding electronic state.

From the constraints in Eqs.~\ref{eq:AHA_constraint_E_u} and~\ref{eq:AHA_contraint_E_d}, we find for the energy of the AHA:
\begin{align}
    E(\lambda, d) = \, &\frac{E_0(d) - E_u}{\lambda_0^2 - \lambda_u^2} \lambda^2 \notag \\ &+ \frac{1}{2} \left( E_0(d) + E_u - \frac{E_0(d) - E_u}{\lambda_0^2 - \lambda_u^2}\left( \lambda_0^2 + \lambda_u^2 \right) \right) \label{eq:E_lambda}
\end{align}

In general, the calibration energy~$E_c$ at~$\lambda_0, d_0$ will be obtained via a quantum chemistry calculation and hence include the information of the corresponding electron density~$\rho_c$. As systems~$A$ and~$B$ are iso-electronic, $\rho_c$ enables access to the first (alchemical) derivative \cite{von_Rudorff_2020, domenichini_2022_al_geom_relax} of the energy w.r.t.~$\lambda$:
\begin{align}
    \label{eq:def_F_0}
    F_c := \frac{\partial E (\lambda)}{\partial \lambda}\Bigg\vert_{\lambda = \lambda_0} = \int_{\mathbb{R}^3} \!\! d\bm{r} \, \rho_c(\bm{r}) \, (v_B - v_A)
\end{align}
Knowledge of~$F_c$ gives a fourth, albeit distance-dependent, constraint on the functional form of~$E(\lambda,d)$. 
With Eq.~\ref{eq:def_F_0} in mind, we could have picked a different triplet of points or derivatives to determine the parabola, e.g.
\begin{align}
    E(\lambda_0, d) &=: E_0(d)\\
    \frac{\partial E(\lambda, d)}{\partial \lambda}\Bigg\vert_{\lambda = \lambda_0} &=: F_0(d)\\
    \frac{\partial E(\lambda, d)}{\partial \lambda}\Bigg\vert_{\lambda = \lambda_m}  &= \,0 \quad \quad ,
\end{align}
which immediately gives:
\begin{align}
    E(\lambda, d) = \frac{F_0(d)}{2\lambda_0} (\lambda^2 - \lambda_0^2) + E_0(d)
\end{align}
Another example is the inclusion of the united atom at only one side of the parabola:
\begin{align}
    E(\lambda_0, d) &=: E_0(d)\\
    \frac{\partial E(\lambda, d)}{\partial \lambda}\Bigg\vert_{\lambda = \lambda_0} &=: F_0(d)\\
    E(+\lambda_u) &=: \,E_u \quad \quad ,
\end{align}
which leads to:
\begin{align}
    \notag
    E(\lambda,d) & = \left[\frac{E_u -E_0(d)}{(\lambda_u - \lambda_0)^2} - \frac{F_0(d)}{(\lambda_u - \lambda_0)}\right] (\lambda-\lambda_0)^2 \\
    & \quad + F_0(d)(\lambda - \lambda_0) + E_0(d)
\end{align}

However, in addition to being another distance-dependent quantity and consequently requiring the calculation of $\partial E_s/\partial \lambda$ (cf.~Sec.~\ref{sec:distance_dep}), both sets of three points rely on the alchemical force~$F_0(d)$, i.e. any implementation necessitates the calculation of Eq.~\ref{eq:E_d}'s derivative w.r.t.~$\lambda$ (see below) in addition to~$F_c$. $F_c$ suffers from the errors made in the construction of basis sets which are optimized on energies rather than densities. 
Thus, for numerical stability, we consider only Eq.~\ref{eq:E_lambda} because its distance dependence can be expressed without derivatives in~$\lambda$.

\subsection{The Alchemical Quartic Approximation (AQuA)}
\label{sec:AQuA}

Going beyond the quadratic approximation, the next higher polynomial in~$E(\lambda)$ must be quartic for symmetry reasons, which we dub the Alchemical Quartic Approximation (AQuA). Using all four previous constraints (Eqs.~\ref{eq:AHA_constraint_E_u}, \ref{eq:AHA_contraint_E_d} and~\ref{eq:def_F_0}), we find for the AQuA:
\begin{align}
    E(\lambda, d) = \,&\left( \frac{F_0(d)}{2\lambda_0 (\lambda_0^2 - \lambda_u^2)} + \frac{E_u - E_0(d)}{(\lambda_0^2 - \lambda_u^2)^2} \right) \lambda^4 \notag \\
    & + \left( -2\lambda_0^2 \frac{E_u - E_0(d)}{(\lambda_0^2 - \lambda_u^2)^2} - \frac{F_0(d)}{2\lambda_0}\frac{\lambda_0^2 + \lambda_u^2}{\lambda_0^2 - \lambda_u^2} \right) \lambda^2 \notag \\
    & + \left( -\lambda_u^4 \frac{E_u - E_0(d)}{(\lambda_0^2 - \lambda_u^2)^2} + \frac{F_0(d)}{2\lambda_0}\frac{\lambda_0^2 \lambda_u^2}{\lambda_0^2 - \lambda_u^2} \right)
\end{align}
However, increasing the polynomial order even further or allowing for fractional powers in~$\lambda$ will exacerbate numerical issues and error propagation from constraints, or even remove the possibility to solve for the coefficients analytically. 
As fifth order polynomials are the highest ones to be analytically solvable, this also excludes more complicated methods like the diatomic energy formula from Ref.~\onlinecite{michael_hammett_JACS} with powers of $\lambda^{7/3}$ from receiving an analytical treatment.

\subsection{Distance-dependence}
\label{sec:distance_dep}

\begin{figure}
    \centering
    \includegraphics[width=\linewidth]{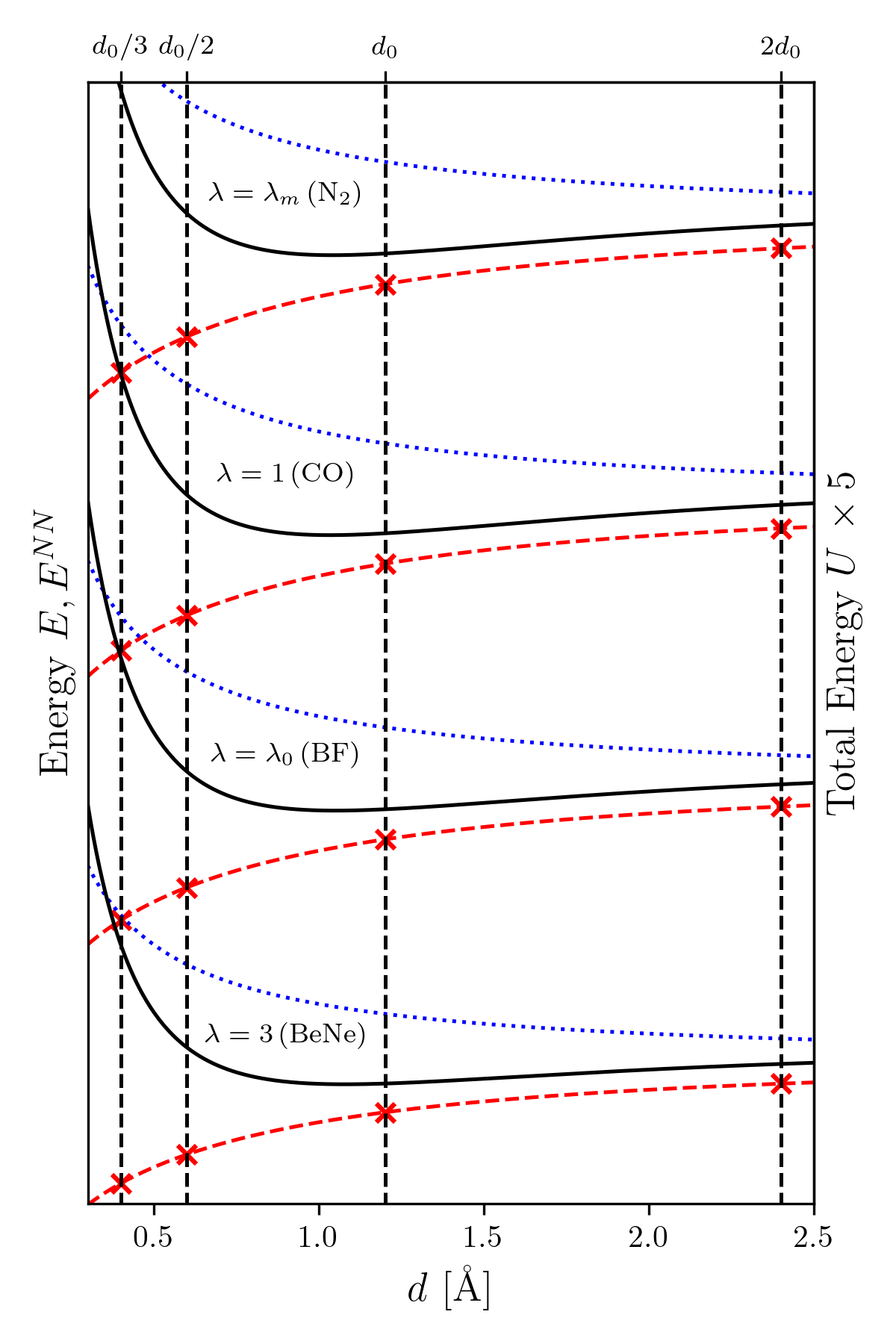}
    \caption{Qualitative decomposition of absolute energy of diatomics as a function of interatomic distance~$d$. Nuclear energy~$E^{NN}$ (dotted, blue lines), electronic energy~$E$ (dashed, red lines, Eq.~\ref{eq:E_d}) and total energy~$U$ (solid, black lines; scaled by factor 5 for clarity). Sum formulas exemplify systems drawn from the 14 electron series. 
    Further annotations are the same as in Fig.~\ref{fig:cartoon_parabola}, i.e. $\lambda_0 = 2$ denotes the calibration diatomic, $\lambda_m = 0$ the peak of the AHA/AQuA. Red markers x connected by black dashed vertical line denote  electronic energies which form the parabola modeled by AHA.
     Energy offsets are for visualization purposes.}
    \label{fig:cartoon_potentials}
\end{figure}

To model the distance-dependence at some fixed $\lambda_0$, we have identified by trial and error the interatomic potential of the diatomic in ground state,
\begin{align}
    E_0 (d) = a + b\, e^{-c\sqrt{d}}
\end{align}
which differs from the attractive (= electronic) part of a Morse potential by the square root in the argument of the exponential function.
To the best of our knowledge, this is a new functional form for modeling the covalent bonding energy of two atoms. 
In the spirit of satisfying extreme close- and far-distance behavior, we immediately find three constraints:
\begin{align}
    E_0 (d_0) &=: E_c \\
    E_0 (0) &= E_u \\
    E_0 (\infty) &= E_1 + E_2 =: E_s
\end{align}
The energies of the neutral atoms $Z_1, Z_2, Z_u$ are neither {$d$-,} nor~$\lambda$-dependent, and can be precomputed (see Computational Details).

Employing the three constraints above, we find:
\begin{align}
    \label{eq:E_d}
    E_0 (d) = (E_u - E_s) \left( \frac{E_c - E_s}{E_u - E_s} \right)^{\sqrt{d/d_0}} + E_s
\end{align}
Implicit in this equation is the physical constraint $E_u < E_c < E_s$, i.e. the calibration must never be the united atom which was already a condition for the AHA/AQuA) but also never the infinitely (or very far) separated system!
With Eq.~\ref{eq:E_d}, only one calibration point $E_c$ at $\lambda_0, d_0$ is needed to determine the energy behavior $E(\lambda, d)$ of iso-electronic diatomics for $\lambda < |\lambda_u|$ and $d \in [0.7,2.5]$\,\AA.

There are two reasons to consider only the electronic distance-dependence: firstly, alchemy only makes relative statements about the electronic part of Schrödinger equations and secondly, as the Coulomb repulsion is known analytically, it appears only natural to model electronic degrees of freedom separately.
However, while Eq.~\ref{eq:E_d} exhibits the correct behavior within the equilibrium range of binding (0.7 - 2.5 \AA), and at infinite distance, the total energy upon addition of Coulomb repulsion does not show yet the desired physical dissociation behavior in the long range.
In order to also treat dissociation, our potential would still need to be morphed into the correct attractive Coulombic and van der Waals dispersion terms which are well known~\cite{HeitlerLondon,EisenschitzLondon}, and continue to be further improved and developed~\cite{khabibrakhmanov2023universal,matej_2024_QDO}.
Furthermore, difficulties are to be expected as the regime of larger interatomic distances is affected by electronic multi-reference effects.
A viualization of different (electronic, nuclear and total) energies of different diatomics and different distances~$d$ can be found in Fig.~\ref{fig:cartoon_potentials}. It serves as complement to Fig.~\ref{fig:cartoon_parabola} such that slices of one graph constitute the functions of the other.

Together with the AHA, our interatomic potentials offers a description of energies in diatomics both in~$\lambda$ and in distance~$d$ from first principles. There are other such approaches e.g. the embedded atom method for crystals of transition metals \cite{finnis_1984,daw_1984} which employs potentials to model the energy in terms of pair-wise distances, and impurities as deviations in nuclear charge from a so-called "host" electron density, albeit in terms of contributions to the electron density itself. Such an ansatz of jellium plus transmutation is reminiscent of a recent prediction of bonding trends in the context of computational alchemy~\cite{michael_hammett_JACS}. In contrast to (semi-)empirical methods, our interatomic potential in combination with the AHA stems purely from the energy constraints.


\section{Numerical results and discussion}

\subsection{Comparing potentials}

\begin{figure}
    \centering
    \includegraphics[width=\linewidth]{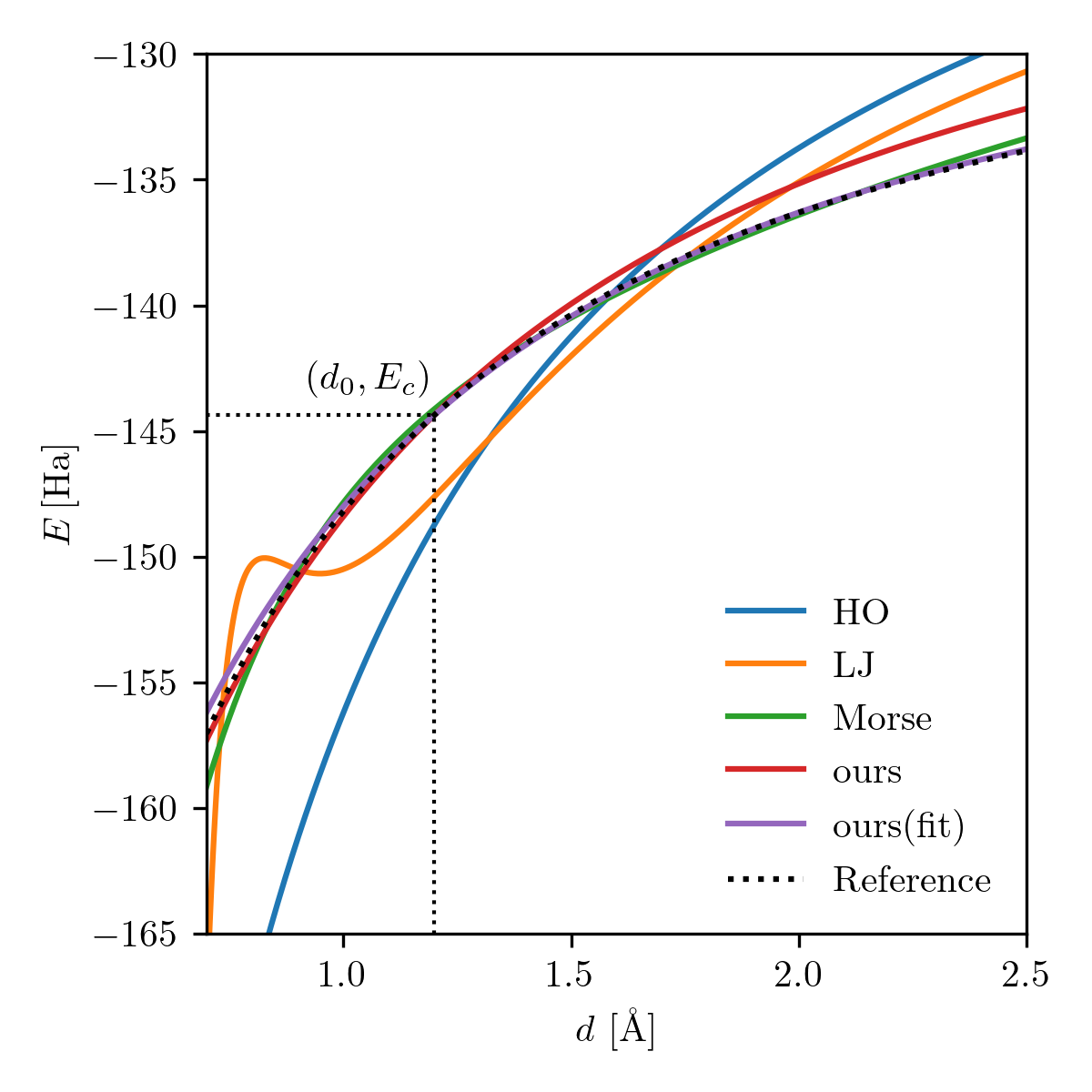}
    \caption{Comparison of various electronic interatomic potentials for BF. Harmonic oscillator (HO), Morse, Lennard-Jones (LJ) (all after subtraction of nuclear Coulomb repulsion), and our potential calibrated to $E_c = E(d_0)$ corresponding to the electronic DFT energy for BF at $d_0 = 1.2${\AA} using either single atom energies $E_s$ pre-computed with DFT (red solid), or treating $E_s$ as another fitted parameter (purple solid). Reference (dotted line) corresponds to DFT, \texttt{pbe0}/\texttt{cc-pVDZ}).}
    \label{fig:potentials}
\end{figure}

\begin{figure}
    \centering
    \includegraphics[width=\linewidth]{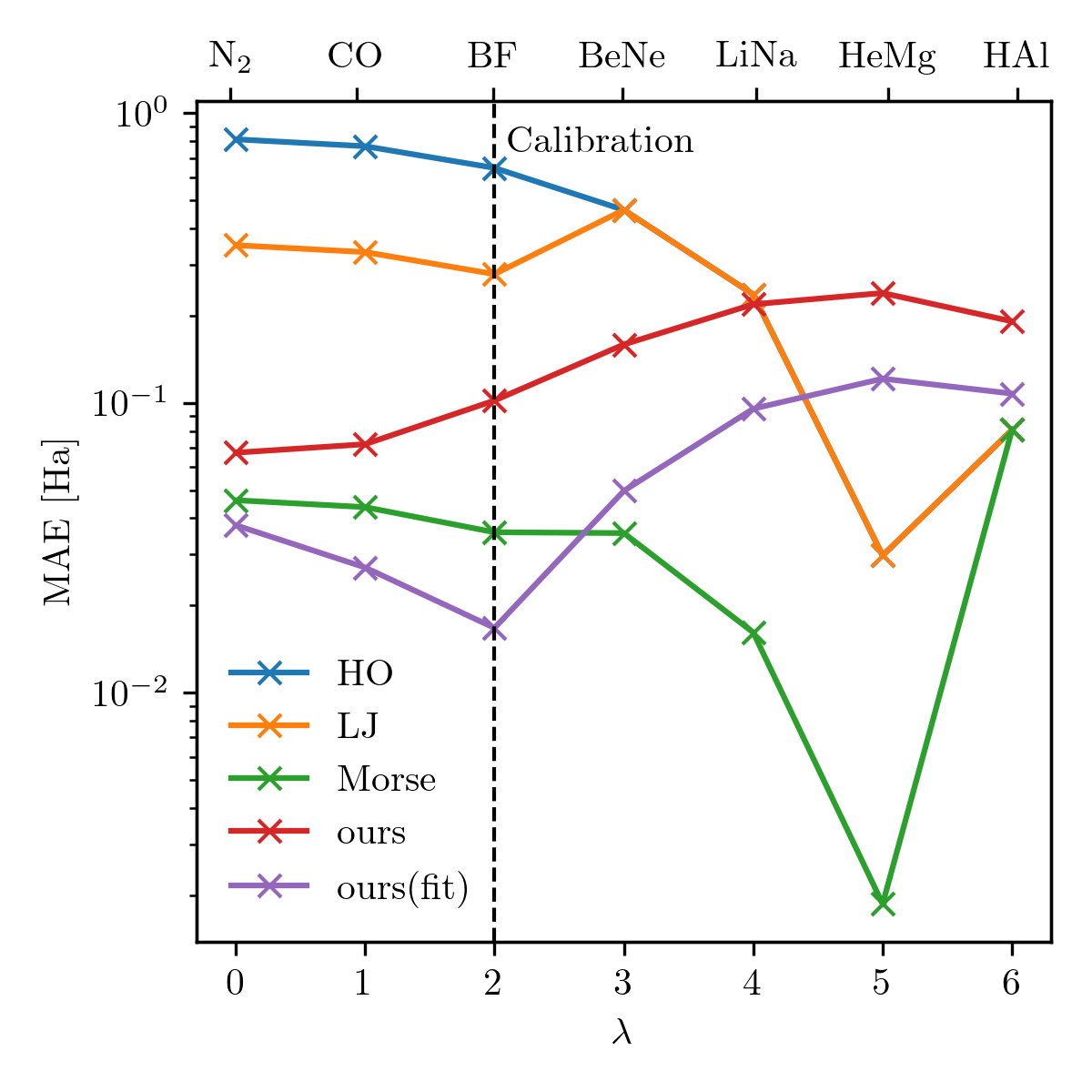}
    \caption{MAE in the AHA model (i.e. $\sum_d | E_{\text{true}} - E(\lambda, d) |$ for 1024 steps $d \in [0.7, 2.5] $ Å, stepsize 0.176\,pm) vs $\lambda$ with BF as calibration. The interatomic potentials are harmonic (HO), Lennard-Jones (LJ), Morse, all with the Coulombic repulsion subtracted, and ours, once with $d_0,E_c,E_u,E_s$ pre-computed (DFT, \texttt{pbe0}/\texttt{cc-pVDZ}), then with $E_s$ treated as a parameter in fit.}
    \label{fig:MAE_BF}
\end{figure}

\begin{figure}
    \centering
    \includegraphics[width=\linewidth]{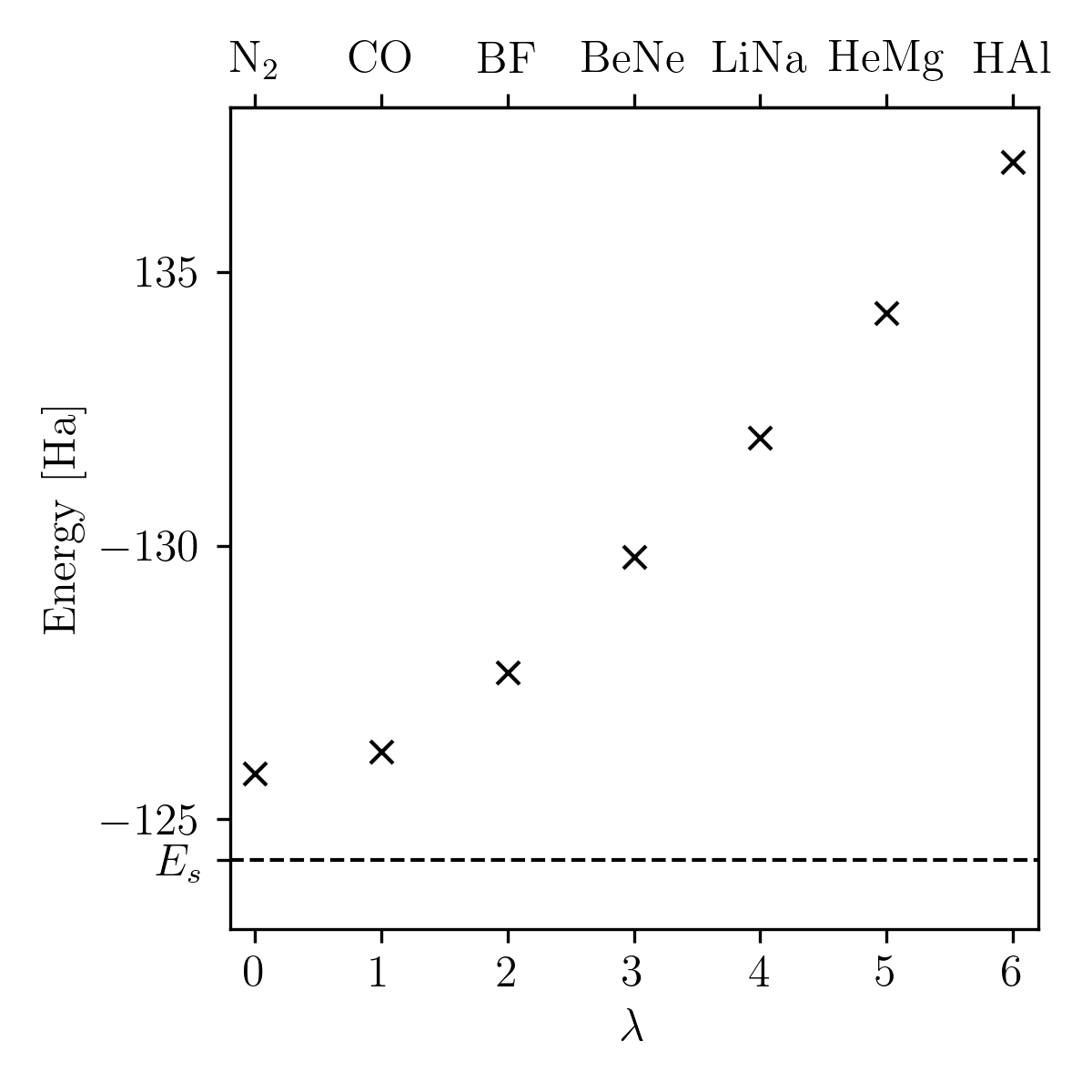}
    \caption{Behavior of $E_s$ when treated as a fitting parameter evaluated at at different systems~$\lambda$ and when precomputed (dashed line). Calibration point is BF ($\lambda_0=2$) at $d_0 = 1.2$\,\AA, $E_u$ is precomputed (DFT, \texttt{pbe0}/\texttt{cc-pVDZ}).}
    \label{fig:E_s_vs_lambda}
\end{figure}

Note that conventional potentials, such as Morse, Lennard-Jones or the harmonic potential (harmonic in~$d$) all model distance-dependent nuclear repulsion \textit{and} electronic attraction together---despite the fact that the Coulombic part could have been easily subtracted in order to better focus on the contributions from quantum mechanics (Fig.~\ref{fig:potentials}). 
Especially for short distances, difficulties arise as Coulombic and empirical repulsion do not cancel exactly, e.g. the repulsive term of the Lennard-Jones (LJ) potential ($\propto r^{-12}$) does not describe a \textit{physical} repulsion. Fitting parameters of the LJ potential minus the Coulomb repulsion to the electronic energy of the diatomic then produces a cancellation of questionable quality as seen in Fig.~\ref{fig:potentials}. 
The advantage of Eq.~\ref{eq:E_d} is its correct and systematic behavior for $d = 0$ and $d \rightarrow \infty$, and its dependence on only one calibration point, almost regardless of choice of $d$ and $\lambda$.

Computing parameters via fitting is not necessary for reasonable MAEs of our potential, but naturally, one can consider the quantity $E_s$ as fitting parameter to improve Eq.~\ref{eq:E_d}'s accuracy. As calibration point, we choose BF ($\lambda_0=2$) and $d_0 = 1.2$ Å.
The different behavior of a precomputed $E_s$ when compared to the fitted one can be found in Fig.~\ref{fig:E_s_vs_lambda}. 
However, gaining improved accuracy in the most relevant range of binding $E_0(d)$, our potential looses part of its first-principle foundation. 
In Fig.~\ref{fig:MAE_BF}, we present a comparison of our potential with established potentials like the harmonic oscillator (HO), Lennard-Jones (LJ) and the Morse potential, all within the AHA.
Again, the calibration point is BF ($\lambda_0=2$) at $d_0 = 1.2$~\AA.
In the fitted potentials (HO, LJ, Morse, our(fit)), there is a clear improvement towards the calibration calculation at $\lambda=2$, but for larger deviations from the symmetric diatomic, the approximation of the AHA worsens (seen in the MAE of ours and ours(fit) which only predict electronic energies). The extreme changes of the other series (HO, LJ, Morse) in MAE stem from a similar source:
an improvement of the MAE for unsymmetric charges $Z_1, Z_2$ can be achieved in the fitting routine until the AHA approximation worsens to such a degree that all three potentials can no longer adequately describe the energy shape. Similar behavior can be found throughout the different calibrations (N$_2$, CO, BeNe, LiNa, HeMg, HAl) and the series’ of 8 (Be$_2$, LiB, HeC, HN), 10 (B$_2$, BeC, LiN, HeO, HF) and 12 (C$_2$,
BN, BeO, LiF, HeNe, HNa) electrons.

\subsection{Machine Learning with AHA as baseline}

\begin{figure*}
    \centering
    \includegraphics[width=0.75\linewidth]{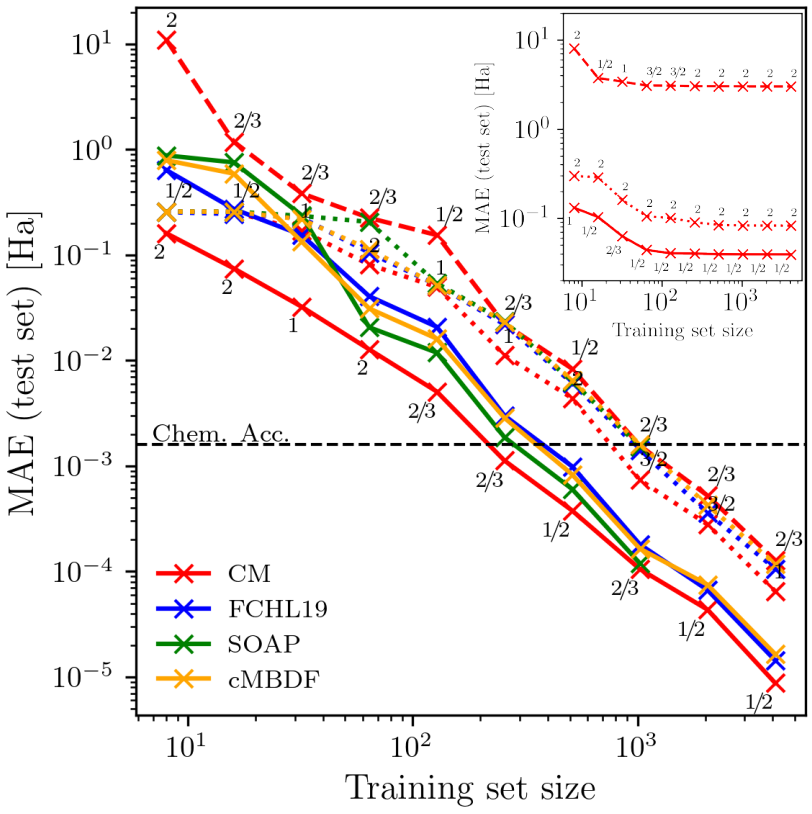}
    \caption{Learning curves of Kernel Ridge Regression (KRR) with calibration system BF. ML models correspond to no baseline model (direct learning, dashed line), with our potential and AHA ($\Delta$-learning, solid line) and the Morse potential and AHA ($\Delta$-learning, dotted line). Exponents $n$ of \texttt{CM(n)} are annotated. Test and training set drawn at random. Inset: Test set exclusively contains CO-samples while training samples are drawn at random from all systems but CO.}
    \label{fig:BF_learning_curves}
\end{figure*}

As a calibration calculation, we again pick BF. Then, the difference between ours and the Morse potential as a baseline model for the prediction of diatomic energies and the reference data at DFT-level of theory (\texttt{pbe0/cc-pVDZ}) can be learned with KRR. We consider DFT to be sufficient for the physical modeling of diatomics while simultaneously retaining low computational cost for the generation of training data. For this, we compute 7 times 1024 points (for the 7 different $\lambda$ and 1024 points for $d \in [0.7, 2.5]$ \, \AA), then choose different global and local representations for the diatomics (see Sec. Computational Details).
The kernel is Laplacian with the Manhattan norm. To determine hyperparameters, we employ 8-fold cross-validation for different training set sizes $N_{\text{train}} = 8,16,32,64 \dots, 4096$. The size of the test set is $N_{\text{test}} = 3072$.
We compare these two $\Delta$-learning approaches with direct learning of the \texttt{CM(n)} representation (Fig.~\ref{fig:BF_learning_curves}).
Note the different sizes of representations: while \texttt{CM(n)} contains only four numbers per diatomic, \texttt{cMBDF} consists of 40 (times 2 atoms), \texttt{FCHL19} of 3952 (times 2 atoms) and finally \texttt{SOAP} with 44296 (times 2 atoms)! Especially \texttt{SOAP} scales costly as every element species of the 14-electron diatomics is included, i.e. elements H to Si.

\begin{figure*}
    \subfloat[]{
        \centering
        \includegraphics[width=0.49\linewidth]{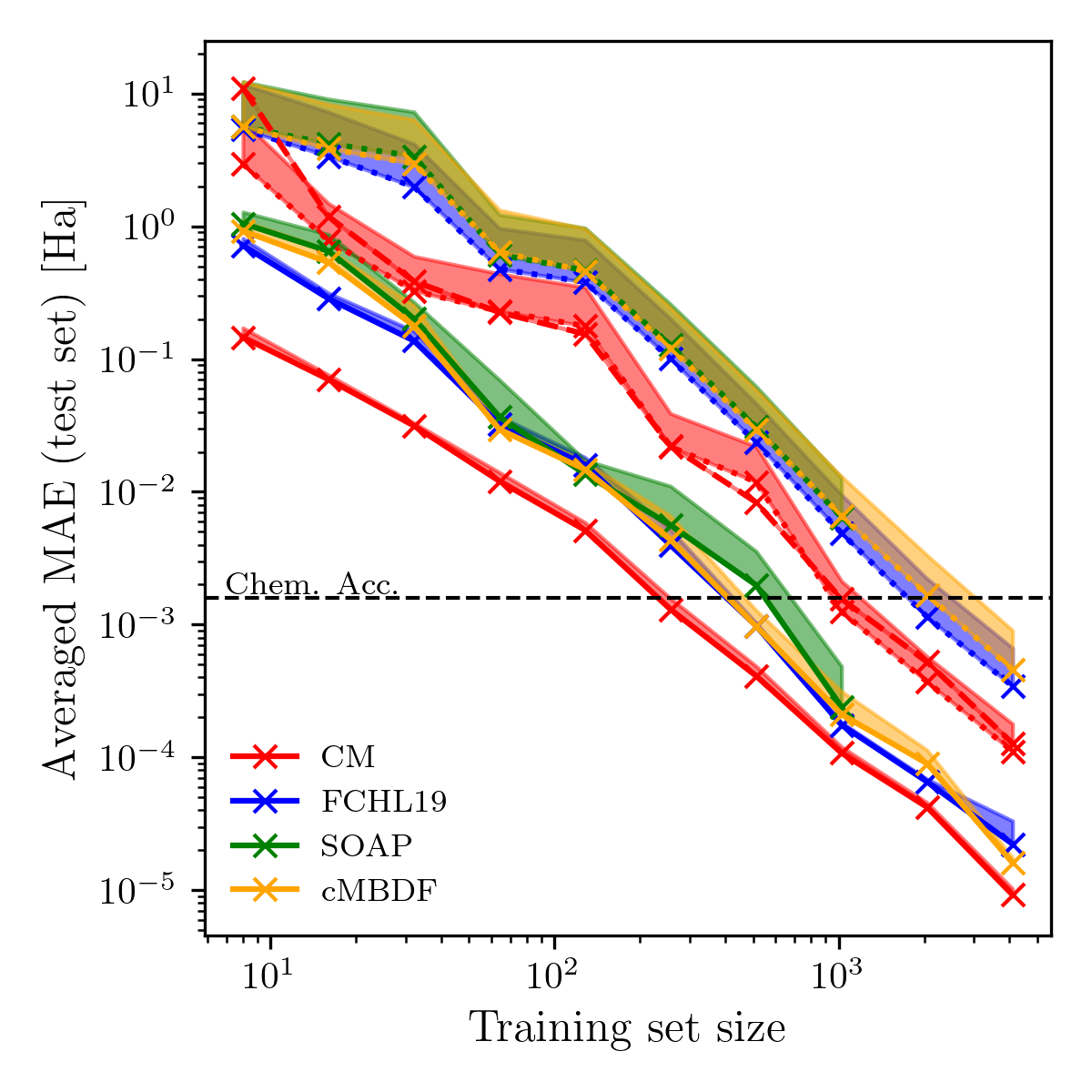}
        \label{fig:N2_series_learning_curves_averaged}
    }
    \subfloat[]{
        \centering
        \includegraphics[width=0.49\linewidth]{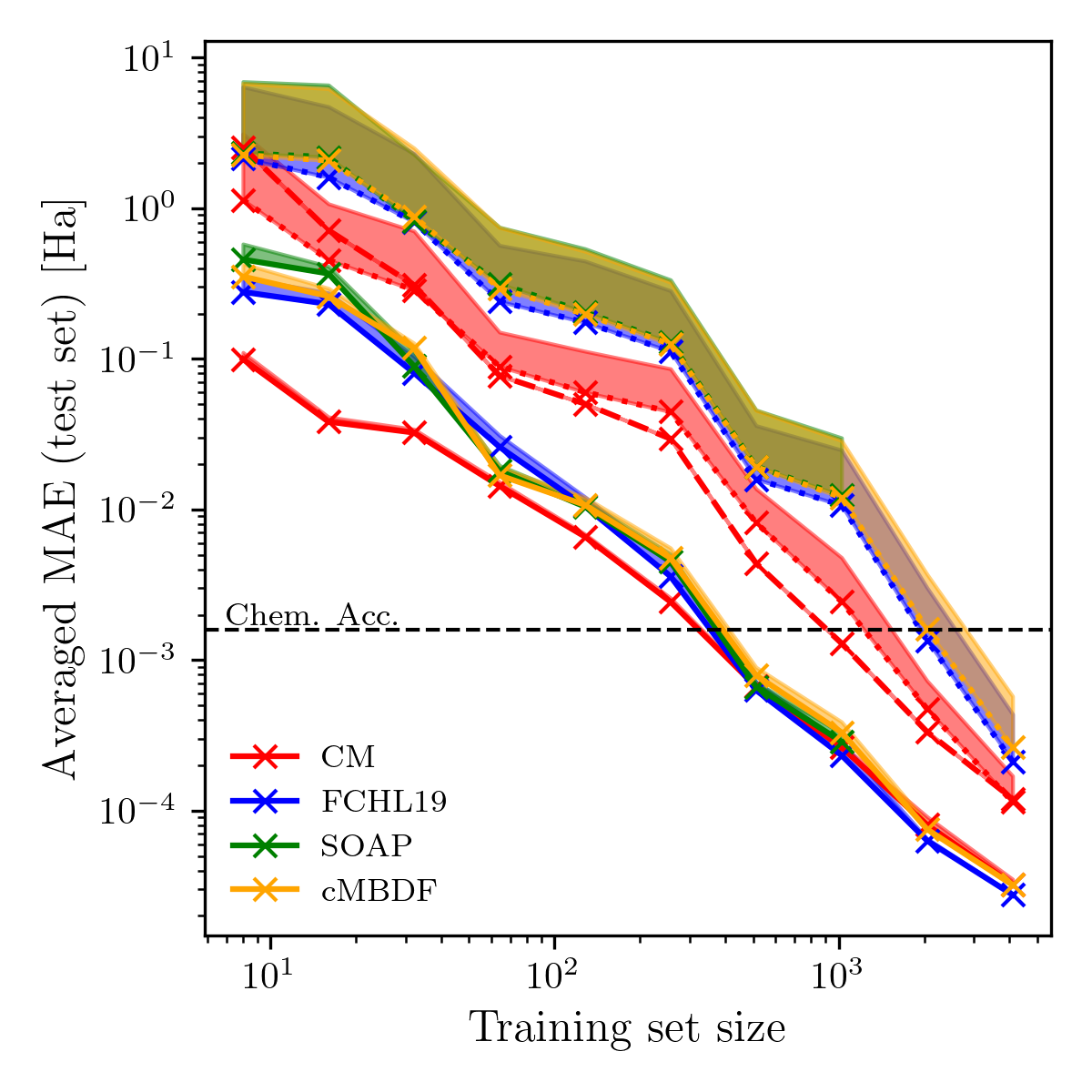}
        \label{fig:C2_series_learning_curves_averaged}
    }\\
    \subfloat[]{
        \centering
        \includegraphics[width=0.49\linewidth]{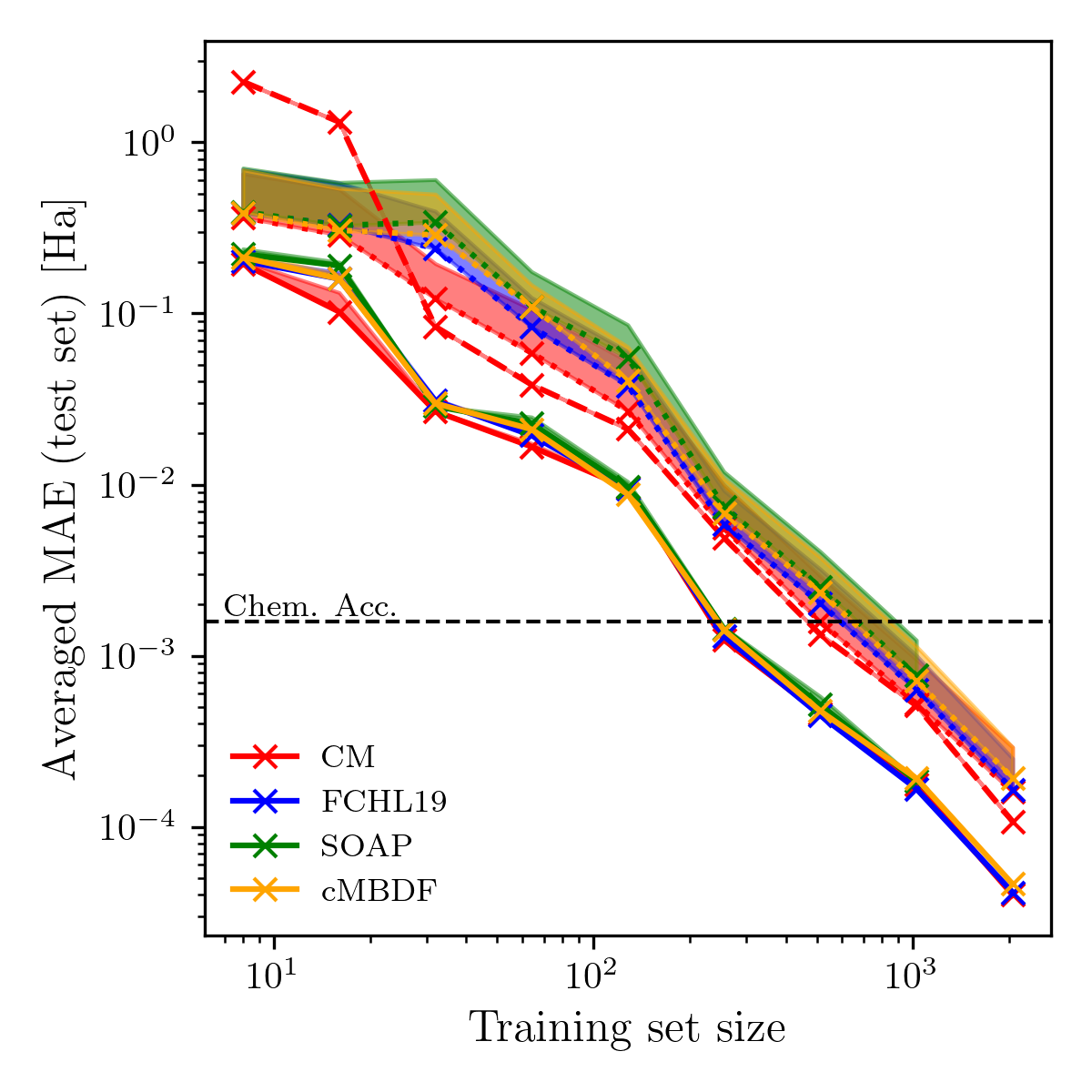}
        \label{fig:B2_series_learning_curves_averaged}
    }
    \subfloat[]{
        \centering
        \includegraphics[width=0.49\linewidth]{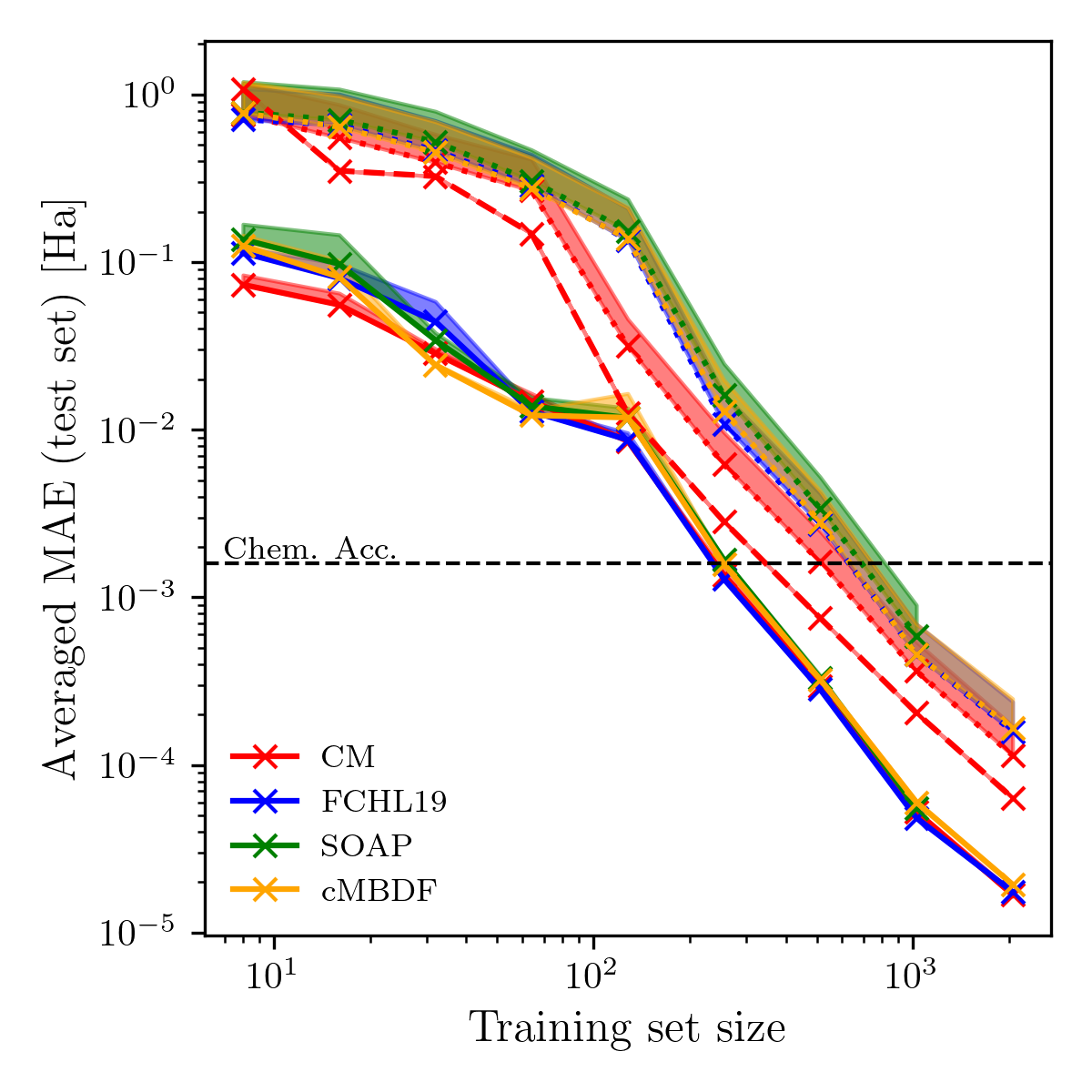}
        \label{fig:Be2_series_learning_curves_averaged}
    }
    \caption{Averaged learning curves of KRR with calibration calculations from all diatomics of the \textbf{(a)} N$_2$-series (N$_2$, CO, BF, BeNe, LiNa, HeMg, HAl), \textbf{(b)} C$_2$-series (C$_2$, BN, BeO, LiF, HeNe, HNa), \textbf{(c)} B$_2$-series (B$_2$, BeC, LiN, HeO, HF), \textbf{(d)} Be$_2$-series (Be$_2$, LiB, HeC, HN).  ML models correspond to no baseline model (direct learning, dashed line), with our potential and AHA ($\Delta$-learning, solid line) and the Morse potential and AHA ($\Delta$-learning, dotted line). Test and training set drawn at random. Shaded areas denote the (positive) standard deviation.
    With decreasing number of electrons, the $\Delta$-learning with a baseline of AHA+Morse proves to be harmful as its MAE increases even beyond direct learning (see (d)).}
    \label{fig:learning_cuves_averaged}
\end{figure*}

In addition, this procedure can be repeated for different calibration systems (N$_2$, CO, BeNe, LiNa, HeMg, HAl) and the iso-electronic series' of 8 (Be$_2$, LiB, HeC, HN), 10 (B$_2$, BeC, LiN, HeO, HF) and 12 (C$_2$, BN, BeO, LiF, HeNe, HNa) electrons, with similar results (Fig.~\ref{fig:learning_cuves_averaged}). Odd numbers of electrons are possible as well, but lead to half-integer $\lambda_m,\lambda_0, \lambda_u$ when considering physical diatomics. However, we did not study these diatomics yet as it would require more sophisticated treatments of the open-shell electron system.

In Fig.~\ref{fig:BF_learning_curves}, the small, global representation~\texttt{CM(n)} clearly works best albeit only marginally. This small difference between global and local kernels diminishes the smaller the parabola of the AHA becomes (N$_2$ to C$_2$ to B$_2$ to Be$_2$, Fig.~\ref{fig:learning_cuves_averaged}). 
This is to be expected since diatomics do not possess three- or higher-order many-body terms and can be adequately described using the interatomic distance information. 

When comparing to direct learning, we appear to gain almost one order of magnitude in accuracy throughout all learning curves from using AHA+ours as a baseline model in $\Delta$-learning, indicating its usefulness as a baseline.

In Fig.~\ref{fig:potentials}, the Morse potential clearly describes the calibration calculation (BF) better; but when testing its generalizability, i.e. its performance when included in the AHA as in Fig.~\ref{fig:BF_learning_curves}, AHA+Morse as baseline model performs worse.
Evidently, AHA+Morse is less systematic when compared to AHA+ours, or put differently, the calibration calculation performed for BF leads to overfitting in the Morse potential which becomes apparent when trying to extrapolate outside of BF. 

Although the Morse potential in itself describes the interatomic behavior between two atoms adequately once parameters are found, these parameters are not derived from physical principles. When paired with the AHA (i.e. a physical model!), its loss of generality becomes obvious. If the diatomic potential in an iso-electronic series produces such problems, a generalization to molecules is even less advised.

When considering the full diatomic series' of N$_2$, C$_2$, B$_2$, Be$_2$ and their different electron numbers, the offset in accuracy of AHA+Morse as baseline in $\Delta$-learning is not just nullified but the baseline itself becomes harmful, i.e. worse than direct learning (cf. Figs~\ref{fig:learning_cuves_averaged}). For this reason, we considered multiple representations (\texttt{CM(n)}, \texttt{FCHL19}, \texttt{SOAP}, \texttt{cMBDF}) to be certain this effect is not a numerical coincidence of one specific representation.

Finally, testing the AQuA model in place of the AHA above results in considerable numerical problems as the MAE of the baseline models (both AQuA+ours and AQuA(fit)+ours vary little) and subsequent $\Delta$-learning with the KRR model decrease for $\lambda \sim \lambda_0$, but drastically increase to thousands of Hartrees upon departing further from~$\lambda_0$. 
We interpret these problems as numerical since a better accuracy was indeed observed close to the calibration but further research will be necessary to determine and possibly remedy the source of these inaccuracies (e.g. computation of $\partial E_s/\partial \lambda$ or $F_0$). 
As basis sets are often optimized w.r.t. minimum energies of one fixed element at a time, and not densities, we hypothesize these issues to originate from basis set errors, previously discussed in Ref.~\onlinecite{giorgio_basisset}.

\subsection{Out-of-sample prediction of diatomics}
\label{sec:outofsample}

We want to determine the MAE of the KRR model when systematically removing one specific diatomic, e.g. CO, from the training set to use as test set. Again, consider the AHA+ours and AHA+Morse baseline models with BF as calibration calculation. Instead of randomly selecting $N_{\text{train}}$ samples, restrict the data in the training set to $\lambda \neq 1$ and in the test set to $\lambda = 1$. Since local representations do not work anymore (cf.~Eq.~\ref{eq:kernel_local}), we exclusively consider the performance between the $\texttt{CM(n)}$ representations in direct and $\Delta$-learning (inset of Fig.~\ref{fig:BF_learning_curves}).
Clearly, all three models approach an average energy prediction early but no accuracy is gained upon further increase of the training set size. 
However, the baseline model of AHA+ours gives a significant edge over the direct learning or $\Delta$-learning with AHA+Morse.
Both the stagnation, as well as the advantage of the physics-based baseline model are to be expected, as the former originates from the lack of samples, while the latter explicitly applies to all iso-electronic diatomics considered here, independent of training set composition.






\section{Conclusion}

We presented the Alchemical Harmonic Approximation (AHA) to describe the iso-electronic series of diatomics relying on only one single calibration point. Going beyond an energy parabola in~$\lambda$ to a fourth order polynomial proved numerically difficult. At the same time, we introduced a new functional form for the electronic potential between two atoms, and compared it to established potentials (HO, LJ, Morse).

It came as no surprise that AHA+Morse outperforms AHA+HO and AHA+LJ, as well as AHA+our potential.
However, after fitting AHA+our potential using $E_s$, it performed better than AHA+Morse for $\lambda \rightarrow \lambda_m$. 
Note that AHA+Morse requires $E_c$ plus 4 parameters for every diatomic (Fig.~\ref{fig:MAE_BF}), while AHA+ours requires 
$E_c$ and only one $E_s$ parameter for every diatomic. 
However, this observation applied only to the calibration system, i.e. one specific diatomic. This is the essence of our findings: outside of the diatomic used in the calibration, our potential proved to be more general because of its derivation from physical principles. 

 AHA+ours and AHA+Morse were used as a baseline for $\Delta$-machine learning based on KRR and commonly used representations (\texttt{CM(n)}, \texttt{FCHL19}, \texttt{SOAP}, \texttt{cMBDF}).
Improvement from direct to $\Delta$-learning (and from AHA+Morse to AHA+ours) shifted learning curves by almost one order of magnitude. This utility, however, depends on a sufficiently diverse training set as demonstrated in the inset of Fig.~\ref{fig:BF_learning_curves}.  
Comparison of the respective performance of AHA+ours and AHA+Morse as baselines suggests that the former reaches chemical accuracy for fewer training instances than the latter. 

For minimally empirical analytical estimates of energetics among diatomics, one might favor AHA+ours as it provides the correct behavior at short and long distances across the iso-electronic chemical space. 
It correctly splits in Coulomb and electronic contributions, and it relies only on one calibration point in the $(\lambda, d)$-surface.


Eq.~\ref{eq:E_d} is sufficient to predict parameters of different potential forms, e.g. the three quantities central to modeling the total energy~$U$ in a Morse potential are the position~$x_{\text{min}}$, depth~$D_{\text{min}}$ and width~$a_{\text{min}}$ of $U$'s minimum. Parameters of more sophisticated potentials like the Expanded Morse Oscillator\cite{meshkov_2014_Be2} or the Morse/Long-range potential and its parameters\cite{leroy_2006_N2,leroy_2009_Li2} become available as well.

Future extensions of this research might deal with (i) inclusion of the interatomic distance dependent alchemical force~$F_0$ into AHA based models, (ii) the extension of chemical space by considering chemical species which are not iso-electronic in total electron count but rather in number of {\em valence} electrons, as already exemplified for ionic crystals~\cite{AlchemyAlisa_2016}, 
covalent bonding~\cite{Samuel-JCP2016}, and band-gaps of (III)-(V) semi-conductors~\cite{Samuel2018bandgaps}, (iii)
extensions to molecules via fragment-based $\Delta$-machine learning~\cite{delta2015} like e.g. atoms-in-molecules (amons) \cite{Amons}, (iv) the
extension of AHA-baseline $\Delta$-learning to multi-level learning~\cite{zaspel2018boosting,heinen2024reducing}, and (v) a generalization of our interatomic potential beyond the ground state. 


\section*{Computational details}
\label{sec:comp_detail}

\subsection{Kernel Ridge Regression (KRR)}

This introduction follows the outline of Refs.~\onlinecite{thesis_lemm} and~\onlinecite{murphy_2012}. 
We seek to find a kernel-based method to map the representations of unseen data $\bm{x}$ from the representation space to a prediction $\hat{y}$ in label space using given data of size $N$, i.e. representations and true labels $(\bm{x}_1, y_1), \dots, (\bm{x}_N, y_N)$. For this, the prediction $\hat{y}$ is given as a linear combination of weighted distances (according to some norm) in kernel space $K(\cdot,\cdot)$ between $\bm{x}$ and all the given input vectors $\bm{x}_1,\dots, \bm{x}_N$:
\begin{align}
    \label{eq:label_prediction}
    \hat{y}(\bm{x}) = \sum_{i = 1}^N \alpha_i \, K(\bm{x},\bm{x}_i)
\end{align}

We wish to minimize the $L_2$-loss of $\hat{y} - y$ w.r.t.~$\bm{\alpha}$, together with $L_2$-Tikhonov regularization to avoid overfitting:
\begin{align}
    \label{eq:L2_loss_plus_reg}
    \text{loss} = \sum_{i = 1}^N \left( \hat{y}(\bm{x}_i) - y_i \right)^2 +\lambda ||\bm{\alpha}||_2^2
\end{align}
$\lambda$ is a hyperparameter weighting the impact of regularization.

Equating the derivative of Eq.~\ref{eq:L2_loss_plus_reg} to zero, we find an analytical solution:
\begin{align}
    \label{eq:KRR_main}
    \bm{\alpha} &= (K + \lambda \, I_{N \times N})^{-1} \, \bm{y} \\
    \label{eq:def_kernelmatrix}
    K_{ij} &= K(\bm{x}_i,\bm{x}_j)
\end{align}
Thus, the training of a KRR model reduces to the computation and subsequent inversion of the kernel matrix~$K$.

Common kernel functions are the Laplacian ($m=1$) and Gaussian functions ($m=2$) with hyperparameter $\sigma$, and common norms inside those functions are the Manhattan ($n=1)$ and Euclidian norm ($n=2)$:
\begin{align}
    K(\bm{x}_i, \bm{x}_j) &= \exp \left( -\frac{||\bm{x}_i - \bm{x}_j||_n^m}{m! \, \sigma^m}\right)
\end{align}
Here, we pick $n=m=1$.

The input vectors~$\bm{x}_i,\bm{x}_j$ discussed above were examples of \textit{global} representations where the compound's entire information will be encoded in one vector without consideration of atomic contributions.
In contrast, a \textit{local} representation of molecule $i$ allows for atom-wise representations, i.e. $\bm{x}_i = \lbrace \bm{x}_i^I, \rbrace$, of its $I$ atoms, and modifies the kernel function to consider only matching nuclear charges~$Z_I, Z_J$:
\begin{align}
    \label{eq:kernel_local}
    K(\bm{x}_i, \bm{x}_j) &= \sum_{I \in i} \sum_{J \in j} \delta_{Z_I, Z_J} \exp \left( -\frac{||\bm{x}_i^I - \bm{x}_j^J||^m_n}{m!\sigma^m}\right)
\end{align}

The representations used in this work the (global) Coulomb matrix\cite{rupp_muller_lilienfeld_2012} with different inverse powers of the atomic distance treated as hyperparameters (\texttt{CM(n)})\cite{CMn}, the convolutional Many-Body Density Functions (\texttt{cMBDF})\cite{khan_2023_mbdf} available on Github under \url{github.com/dkhan42/cMBDF} with \texttt{rstep=1e-6}, the (local) Smooth Overlap of Atomic Positions (\texttt{SOAP})\cite{soap} implemented in \texttt{DScribe}\cite{dscribe}, and the (local) Faber-Christensen-Huang-Lilienfeld representation from 2019 (\texttt{FCHL19})\cite{fchl_1,fchl_2} implemented in the QML code\cite{qmlcode}. The representation were not optimized to diatomics, but mainly used "as is".

To evaluate their performance, the available data and labels are randomly split into training and test sets. Training sets of size $N_{\text{train}}$ produce a kernel matrix whose accuracy is assessed via the mean absolute error (MAE) upon prediction of labels in the test set:
\begin{align}
    \text{MAE} &= \sum_{j = 1}^{N_{\text{test}}} || y_j -  \hat{y}(\bm{x}_j) ||_1 \\
    &= \sum_{j = 1}^{N_{\text{test}}} \Big|\Big| y_j - \sum_{i = 1}^{N_{\text{train}}} \alpha_i \, K(\bm{x}_j,\bm{x}_i)\Big|\Big|_1
\end{align}
The learning curves below show either this MAE against~$N_{\text{train}}$, or the average and standard deviation of multiple models' MAE against~$N_{\text{train}}$ for cases where models of equal electron number are pooled together (cf.~Fig.~\ref{fig:learning_cuves_averaged} in case of 14, 12, 10, and 8 electrons).

The hyperparameters $\sigma, \lambda$ (and $n$ when considering \texttt{CM(n)}) are determined via 8-fold cross-validation, i.e. 8-fold splitting of the training set in training and validation sets to determine the best set of hyperparameters without training on the test set.

There has been no discussion so far about the labels used in the model. If the labels (energies) of a given input (compound) are known to desired accuracy, one might argue against an intermediate model for the prediction of energy labels as in Sec.~\ref{sec:AHA}, and instead predict the correct labels directly. This is possible but at the cost of data efficiency (cf.~Fig.~\ref{fig:BF_learning_curves}); learning the difference to a baseline model (derived from physical insights) saves training data ($\Delta$-learning~\cite{delta2015}). 
As physical insight is the desired quantity anyhow, comparing $\Delta-$ to direct learning can also serve as a metric for the quality of our baseline model.

\subsection{Software}
Software for the purpose of data generation (e.g. quantum chemistry software) are provided by the \texttt{Python}-packages \texttt{PySCF} \cite{PySCF1,PySCF2} and specifically its internal function \texttt{dft.ks}, \texttt{basissetexchange} \cite{bse1,bse2,bse3}, \texttt{NumPy} \cite{numpy} and \texttt{SciPy} \cite{scipy}. Visualizations were created using \texttt{Matplotlib} \cite{matplotlib}.

\subsection{Data and code availability}

The code that produces the figures and findings of this study, specifically the scripts for generation of DFT data, in addition to the data itself, are openly available on Zenodo under \url{www.zenodo.org/records/13844083}.

Below we give the energies of elements H to Si employed for the generation of data in AHA+ours. Their spin has been kept either zero or one, depending on even or odd number of electrons, respectively. A minimal working example for $Z=3$ with Kohn-Sham-DFT, basis set \texttt{cc-pVDZ} \cite{ccpvnz1,ccpvnz2,ccpvnz3} and the \texttt{pbe0}-functional \cite{PBE0,PBE01,PBE02} using \texttt{PySCF} \cite{PySCF1,PySCF2} reads:

\begin{verbatim}
from pyscf import gto, dft
import basis_set_exchange as bse

Z = 3
basis_set = 'cc-pVDZ'
xc = 'pbe0'

basis = bse.get_basis(basis_set, fmt='nwchem')
mol = gto.M(atom = str(Z)+' 0 0 0',
            charge=0,
            spin=Z%2,
            basis = basis)

mf = dft.KS(mol)
mf.xc = xc
mf.kernel()
\end{verbatim}

\noindent
This results in the numerical values in Tab.~\ref{tab:E_i}.
\begin{table}
    \centering
    \begin{tabular}{c||rcc}
        Nuclear charge $Z$ & Energy [Ha] \\
        \hline
        \hline
         0 & 0.0000 && \\
         1 & -0.5002 && \\
         2 & -2.8876 &&\\
         3 & -7.4661 &&\\
         4 & -14.6351 &&\\
         5 & -24.6159 &&\\
         6 & -37.7241 &&\\
         7 & -54.4864 &&\\
         8 & -74.8817 &&\\
         9 & -99.6356&&\\
         10 & -128.8050&& \\
         11 & -162.1752 &&\\
         12 & -199.9610 &&\\
         13 & -242.2462 &&\\
         14 & -289.1975 &&\\
    \end{tabular}
    \caption{Energies of elements H to Si employed for the generation of data with AHA+ours}
    \label{tab:E_i}
\end{table}

\section*{Supplementary Material}

This article includes no supplementary material.

\section*{Acknowledgements}
We acknowledge discussions with Florian Bley, Oliver Eberle and Katharine Hunt.
We acknowledge the support of the Natural Sciences and Engineering Research Council of Canada (NSERC), [funding reference number RGPIN-2023-04853]. Cette recherche a été financée par le Conseil de recherches en sciences naturelles et en génie du Canada (CRSNG), [numéro de référence RGPIN-2023-04853].
This research was undertaken thanks in part to funding provided to the University of Toronto's Acceleration Consortium from the Canada First Research Excellence Fund,
grant number: CFREF-2022-00042.
O.A.v.L. has received support as the Ed Clark Chair of Advanced Materials and as a Canada CIFAR AI Chair.
O.A.v.L. has received funding from the European Research Council (ERC) under the European Union’s Horizon 2020 research and innovation programme (grant agreement No. 772834).

\section*{Author Contributions}
\textbf{Simon León Krug}: conceptualization (equal), data curation (lead), formal analysis (lead), investigation (equal), methodology (equal), software, visualization (equal), writing - original draft (lead), writing - review \& editing (equal).
\textbf{Danish Khan}: formal analysis (supporting), data curation (supporting), software, visualization (equal), writing - original draft (supporting).
\textbf{O.~Anatole von~Lilienfeld}: conceptualization (equal), formal analysis (supporting), investigation (equal), methodology (equal), funding acquisition, project administration, resources, supervision (lead), visualization (equal), writing - review \& editing (equal).

All authors read and approved the final manuscript.

\section*{Conflict of Interest}
The authors have no conflicts to disclose.

\bibliography{refs.bib}{}
\bibliographystyle{unsrt}

\end{document}